\documentclass[aps,twocolumn,showpacs,nofootinbib]{revtex4}

\usepackage{graphicx}
\usepackage{amsmath}
\usepackage{amssymb}

\newcommand{\vv}[1]{{\bf{#1}}}

\begin{document}

\title{Doping dependence of electromagnetic response in electron-doped cuprate superconductors}

\author{Zheyu Huang, Huaisong Zhao, and Shiping Feng}
\affiliation{Department of Physics, Beijing Normal University,
Beijing 100875, China}


\begin{abstract}
Within the framework of the kinetic energy driven superconducting mechanism, the doping dependence of the electromagnetic response in the
electron-doped cuprate superconductors is studied. It is shown that although there is an electron-hole asymmetry in the phase diagram, the
electromagnetic response in the electron-doped cuprate superconductors is similar to that observed in the hole-doped cuprate superconductors.
The superfluid density depends linearly on temperature, except for the strong deviation from the linear characteristics at the extremely low
temperatures.
\end{abstract}


\maketitle

The parent compounds of cuprate superconductors are Mott insulators with an antiferromagnetic (AF) long-range order (AFLRO) \cite{Kastner98},
then the AF phase is subsided and superconductivity is realized by doping a moderate amount of holes or electrons into these Mott insulators
\cite{bednorz86,tokura89}. It has been found that only an approximate symmetry in the phase diagram exists about the zero doping line between
the hole-doped and electron-doped cuprate superconductors \cite{damascelli03,Krockenberger08}, and the significantly different behaviors of
the hole-doped and electron-doped cuprate superconductors are observed, reflecting the electron-hole asymmetry. Among the various unusual
properties in cuprate superconductors, the electromagnetic response is a key ingredient to understand the still unresolved mechanism of
superconductivity \cite {bonn96}. To elucidate it, the knowledge of the magnetic field penetration depth (then the superfluid density) and
its evolution with doping and temperature is of crucial importance \cite {bonn96}. Experimentally, by virtue of systematic studies using
the muon-spin-rotation measurement technique, some essential features of the electromagnetic response in cuprate superconductors have been
established now for all the temperature $T\leq T_{c}$ throughout the SC dome. For the hole-doped cuprate superconductors
\cite{bonn96,suter04,kamal98,bernhard01,Broun07}, an agreement for the electromagnetic response has emerged that a simple d-wave
superconducting (SC) gap leads to a crossover of the superfluid density from the linear temperature dependence at low temperatures to a
nonlinear one at the extremely low temperatures. However, there are some controversies in the electron-doped side. The early muon-spin-rotation
experimental results of the electron-doped cuprate superconductors \cite{Wu93,Andreone94} showed that the electrodynamics are consistent with a
gaped s-wave behavior. Later, the muon-spin-rotation experimental data \cite{Kim03,Skinta02,Cooper96} showed that at the low doping levels, the
superfluid density at low temperatures is quadratic in temperature, but at the higher dopings, the superfluid density has an activated behavior,
suggesting a d-wave to s-wave pairing transition near the optimal doping. However, the recent muon-spin-rotation experimental results
\cite{Snezhko04,Kokales00,Prozorov00} showed that the superfluid density exhibits a linear temperature behavior, indicative of a pure d-wave
state. In particular, these experimental results \cite{Snezhko04,Kokales00,Prozorov00} found a behavior of the electromagnetic response of
the electron-doped cuprate superconductors similar to that observed in the hole-doped case \cite{bonn96,suter04,kamal98,bernhard01,Broun07}.
Theoretically, the most of the interpretations for the unusual electromagnetic response are focused on the hole-doped cuprate superconductors.
In order to elucidate the mechanism of superconductivity, it is necessary to look into electron-doped cuprate superconductors too, and then to
identify the differences and similarities between the hole-doped and electron-doped cuprate superconductors.

In our recent work \cite{feng10}, the electromagnetic response in the hole-doped cuprate superconductors has been studied based on the kinetic
energy driven SC mechanism \cite{feng0306,feng08}, and then the main features of the doping and temperature dependence of the local magnetic
field profile, the magnetic field penetration depth, and the superfluid density observed on the hole-doped cuprate superconductors
\cite{bonn96,suter04,kamal98,bernhard01,Broun07} are well reproduced. In this paper, we study the doping and temperature dependence of the
electromagnetic response in the electron-doped cuprate superconductors along with this line. We show explicitly that although the electron-hole
asymmetry is observed in the phase diagram \cite{damascelli03,Krockenberger08}, the main features of the electromagnetic response in the
electron-doped cuprate superconductors are similar to that observed in the hole-doped cuprate superconductors \cite{feng10}. The superfluid
density is the wide range of linear temperature dependence at low temperature, extending from close to the SC transition temperature to down to
the temperatures $T\sim 8$K for different doping concentrations, then at the extremely low temperatures $T< 8$K, the superfluid density crosses
over to a nonlinear temperature behavior.

It is commonly accepted that the essential physics of cuprate superconductors is properly accounted by the two-dimensional $t$-$J$ model on a
square lattice \cite{anderson87}. This $t$-$J$ model with the nearest neighbor hopping $t$ has a particle-hole symmetry because the sign of $t$
can be absorbed by changing the sign of the orbital on one sublattice. However, the particle-hole asymmetry in the phase diagram of cuprate
superconductors can be described by including the next-nearest neighbor hopping $t'$ \cite{hybertson90,gooding94,pavarini01}, which therefore
plays an important role in explaining the difference between electron and hole doping. Furthermore, for discussions of the doping and temperature
dependence of the electromagnetic response in the electron-doped cuprate superconductors, the $t$-$t'$-$J$ model can be extended by including the
exponential Peierls factors as,
\begin{eqnarray}\label{t-jmodel}
H&=&t\sum_{i\hat{\eta}\sigma}e^{-i({e}/{\hbar})\vv{A}(l)\cdot\hat{\eta}}PC^{\dag}_{i\sigma}C_{i+\hat{\eta}\sigma}P^{\dag}\nonumber\\
&-&t'\sum_{i\hat{\tau}\sigma}e^{-i({e}/{\hbar})\vv{A}(l)\cdot\hat{\tau}}PC^{\dag}_{i\sigma}C_{i+\hat{\tau}\sigma}P^{\dag}\nonumber\\
&-&\mu\sum_{i\sigma} PC^{\dag}_{i\sigma}C_{i\sigma}P^{\dag}+J\sum_{i\hat{\eta}}{\bf S}_i\cdot{\bf S}_{i+\hat{\eta}},
\end{eqnarray}
where $t<0$, $t'<0$, $\hat{\eta}=\pm\hat{x},\pm\hat{y}$, $\hat{\tau}=\pm\hat{x} \pm\hat{y}$, $C^{\dagger}_{i\sigma}$ ($C_{i\sigma}$) is the
electron creation (annihilation) operator, ${\bf S}_{i}=(S_{i}^{x},S_{i}^{y},S_{i}^{z})$ are spin operators, and $\mu$ is the chemical potential.
In the Hamiltonian (\ref{t-jmodel}), the hopping terms together with exponential Peierls factors account for the coupling of the electron charge
to an external magnetic field \cite{feng10} in terms of the vector potential $\vv{A}(l)$, while the nontrivial part resides in the projection
operator $P$ which restricts the Hilbert space to exclude the zero occupancy in the electron-doped cuprate superconductors, i.e.,
$\sum_{\sigma}C^{\dagger}_{i\sigma}C_{i\sigma}\geq 1$.

For description of the hole-doped cuprate superconductors, the charge-spin separation (CSS) fermion-spin theory \cite{feng04,feng08} has been
developed for incorporating the electron single occupancy local constraint. However, to apply this CSS fermion-spin theory to the electron-doped
counterparts, the Hamiltonian (\ref{t-jmodel}) should be rewritten in terms of a particle-hole transformation
$C_{i\sigma}\rightarrow f^{\dagger}_{i-\sigma}$ as \cite{cheng07},
\begin{eqnarray}\label{t-jmodel1}
H&=&-t\sum_{i\hat{\eta}\sigma}e^{-i({e}/{\hbar})\vv{A}(l)\cdot\hat{\eta}}f^{\dag}_{i\sigma}f_{i+\hat{\eta}\sigma}\nonumber\\
&+&t'\sum_{i\hat{\tau}\sigma}e^{-i({e}/{\hbar})\vv{A}(l)\cdot\hat{\tau}}f^{\dag}_{i\sigma}f_{i+\hat{\tau}\sigma}\nonumber\\
&+&\mu\sum_{i\sigma} f^{\dag}_{i\sigma}f_{i\sigma}+J\sum_{i\hat{\eta}}{\bf S}_i\cdot{\bf S}_{i+\hat{\eta}},
\end{eqnarray}
then the local constraint without the zero occupancy $\sum_{\sigma}C^{\dagger}_{i\sigma}C_{i\sigma}\geq 1$ is transferred as the single occupancy
local constraint $\sum_{\sigma}f^{\dagger}_{i\sigma}f_{i\sigma}\leq 1$, where $f^{\dagger}_{i\sigma}$ ($f_{i\sigma}$) is the hole creation
(annihilation) operator. Now we follow the CSS fermion-spin theory \cite{feng04,feng08}, and decouple the hole operators as
$f_{i\uparrow}=a^{\dagger}_{i\uparrow}S^{-}_{i}$ and $f_{i\downarrow}=a^{\dagger}_{i\downarrow}S^{+}_{i}$, respectively, where the spinful fermion
operator $a_{i\sigma}=e^{-i\Phi_{i\sigma}}a_{i}$ represents the charge degree of freedom together with some effects of the spin configuration
rearrangements due to the presence of the doped electron itself (charge carrier), while the spin operator $S_{i}$ represents the spin degree of
freedom, then the single occupancy local constraint in the electron-doped cuprate superconductors in the hole representation is satisfied in
analytical calculations. In this CSS fermion-spin representation, the Hamiltonian (\ref{t-jmodel1}) can be expressed as,
\begin{widetext}
\begin{eqnarray}\label{t-jmodel2}
H&=&t\sum_{i\hat{\eta}}e^{-i({e}/{\hbar})\vv{A}(l)\cdot\hat{\eta}}(a^{\dagger}_{i+\hat{\eta}\uparrow}a_{i\uparrow}S^{+}_{i}
S^{-}_{i+\hat{\eta}}+a^{\dagger}_{i+\hat{\eta}\downarrow}a_{i\downarrow}S^{-}_{i}S^{+}_{i+\hat{\eta}})
-t'\sum_{i\hat{\tau}}e^{-i({e}/{\hbar})\vv{A}(l)\cdot\hat{\tau}}(a^{\dagger}_{i+\hat{\tau}\uparrow}a_{i\uparrow}S^{+}_{i}
S^{-}_{i+\hat{\tau}}+a^{\dagger}_{i+\hat{\tau}\downarrow}a_{i\downarrow}S^{-}_{i}S^{+}_{i+\hat{\tau}}) \nonumber \\
&-&\mu\sum_{i\sigma}a^{\dagger}_{i\sigma}a_{i\sigma}+J_{{\rm eff}}\sum_{i\hat{\eta}}{\bf S}_{i}\cdot {\bf S}_{i+\hat{\eta}},
\end{eqnarray}
\end{widetext}
where $J_{\rm {eff}}=(1-\delta)^2J$, and $\delta=\langle a^{\dag}_{i\sigma}a_{i\sigma}\rangle=\langle a^{\dag}_{i}a_{i}\rangle$ is the electron
doping concentration.

As in the discussions of the hole-doped case \cite{feng0306}, the SC order parameter for the electron Cooper pair in the electron-doped cuprate
superconductors also can be defined as,
\begin{eqnarray}
\Delta &=&\langle C^{\dag}_{i\uparrow}C^{\dag}_{j\downarrow}-C^{\dag}_{i\downarrow}C^{\dag}_{j\uparrow}\rangle=\langle a_{i\uparrow}
a_{j\downarrow}S^{\dag}_{i}S^{-}_{j}-a_{i\downarrow}a_{j\uparrow}S^{-}_{i}S^{+}_{j}\rangle \nonumber\\
&=&-\langle S^{+}_{i}S^{-}_{j} \rangle\Delta_{a},
\end{eqnarray}
with the charge carrier pairing gap parameter $\Delta_{a}=\langle a_{j\downarrow} a_{i\uparrow}-a_{j\uparrow}a_{i\downarrow}\rangle$. For the
hole-doped cuprate superconductors, a large body of experimental data \cite{damascelli03,ding96} indicate that the hot spots are located close
to the antinodal points of the Brillouin zone, resulting in a monotonic (simple) d-wave gap. However, in contrast, a number of experiments,
including angular resolved photoemission \cite{sato01,matsui05}, scanning SC quantum interference device measurements \cite{tsuei00}, Raman
scattering \cite{blumberg02}, and phase sensitive study \cite{tomaschko12} show that the hot spots are located much closer to the zone diagonal
in the electron-doped side, leading to a nonmonotonic d-wave gap,
\begin{eqnarray}\label{gap}
\Delta({\bf k})=\Delta[\gamma^{(d)}_{{\bf k}}-B\gamma^{(2d)}_{{\bf k }}],
\end{eqnarray}
where $\gamma^{(d)}_{{\bf k}}=[{\rm cos}k_{x}-{\rm cos}k_{y}]/2$ and $\gamma^{(2d)}_{{\bf k}}=[{\rm cos}(2k_{x})-{\rm cos}(2k_{y})]/2$, then
the maximum gap is observed not at the nodal points as expected from the monotonic d-wave gap, but at the hot spot between nodal and antinodal
points, where the AF spin fluctuation most strongly couples to electrons, supporting a spin-mediated pairing mechanism.

In the case of zero magnetic field, it has been shown \cite{feng0306,feng08} in terms of Eliashberg's strong coupling theory that the charge
carrier-spin interaction from the kinetic energy term in the Hamiltonian (\ref{t-jmodel2}) induces a charge carrier pairing state with the
d-wave symmetry by exchanging spin excitations, then the SC transition temperature is identical to the charge carrier pair transition
temperature. Moreover, this d-wave SC state is controlled by both SC gap function and quasiparticle coherence, which leads to that the maximal
SC transition temperature occurs around the optimal doping, and then decreases in both underdoped and overdoped regimes. In particular, within
this kinetic energy driven SC mechanism \cite{guo06,feng08}, some main features of the doping dependence of the low-energy electronic structure
\cite{cheng07}, the dynamical spin response \cite{cheng08}, and the electronic Raman response \cite{geng11} in the electron-doped cuprate
superconductors have been quantitatively reproduced. Following these previous discussions \cite{cheng07,cheng08,geng11}, the full charge carrier
Green function of the electron-doped cuprate superconductors can be obtained in the Nambu representation as,
\begin{eqnarray}\label{holegreenfunction}
g({\bf{k}},i\omega_n)=Z_{\rm{aF}}\,\frac{i\omega_{n}\tau_{0}+\bar{\xi}_{{\rm a}\bf{k}}\tau_{3}-\bar{\Delta}_{\rm{aZ}}({\bf{k}})\tau_{1}}
{(i\omega_n)^{2}-E_{{\rm{a}}{\bf{k}}}^{2}},
\end{eqnarray}
where $\tau_{0}$ is the unit matrix, $\tau_{1}$ and $\tau_{3}$ are Pauli matrices, the renormalized charge carrier excitation spectrum
$\bar{\xi}_{{\rm a}\bf k}=Z_{\rm aF}\xi_{{\rm a}\bf k}$, with the mean-field charge carrier excitation spectrum
$\xi_{{\rm a}\bf k}=Zt\chi_{1} \gamma^{(s)}_{{\bf k}}-Zt'\chi_{2}\gamma^{(2s)}_{{\bf k}}-\mu$, the spin correlation functions
$\chi_{1}=\langle S_{i}^{+}S_{i+\hat{\eta}}^{-} \rangle$ and $\chi_{2}= \langle S_{i}^{+}S_{i+\hat{\tau}}^{-}\rangle$,
$\gamma^{(s)}_{\bf k}=(1/Z)\sum_{\hat{\eta}}e^{i{\bf k}\cdot\hat{\eta}}$,
$\gamma^{(2s)}_{\bf k}= (1/Z)\sum_{\hat{\tau}}e^{i{\bf k}\cdot\hat{\tau}}$, $Z$ is the number of the nearest neighbor or next-nearest neighbor
sites, the renormalized charge carrier d-wave pair gap $\bar{\Delta}_{\rm aZ} ({\bf k})=Z_{\rm aF}\bar{\Delta}_{\rm a}({\bf k})$, with the
effective charge carrier d-wave pair gap $\bar{\Delta}_{\rm a}({\bf k})=\bar{\Delta}_{\rm a}[\gamma^{(d)}_{{\bf k}}-B\gamma^{(2d)}_{{\bf k }}]$,
and the charge carrier quasiparticle spectrum $E_{{\rm a}{\bf k}}=\sqrt{\bar{\xi}^{2}_{{\rm a}{\bf k}}+|\bar{\Delta}_{\rm aZ}({\bf k} ) |^{2}}$,
while the effective charge carrier pair gap $\bar{\Delta}_{\rm a}({\bf k})$ and the quasiparticle coherent weight $Z_{\rm aF}$ satisfy the
following equations \cite{cheng07} $\bar{\Delta}_{\rm a}({\bf k})=\Sigma^{(a)}_{2}({\bf k},\omega=0)$ and
$Z^{-1}_{\rm aF}=1-\Sigma^{(a)}_{1{\rm o}}({\bf k},\omega=0)\mid_{{\bf k}=[\pi,0]}$, where $\Sigma^{(a)}_{1}({\bf k},\omega)$ and
$\Sigma^{(a)}_{2}({\bf k},\omega)$ are the charge carrier self-energies obtained from the spin bubble in the charge carrier particle-hole and
particle-particle channels, respectively, and have been given in Ref. \onlinecite{cheng07} except the effective charge carrier monotonic d-wave
gap has been replaced by the present nonmonotonic one, while $\Sigma^{(a)}_{1{\rm o}}({\bf k},\omega)$ is the antisymmetric part of
$\Sigma^{(a)}_{1}({\bf k},\omega)$. These equations have been solved simultaneously with other self-consistent equations \cite{cheng07,cheng08},
then all order parameters and chemical potential $\mu$ have been determined by the self-consistent calculation.

For discussions of the electromagnetic response in the electron-doped cuprate superconductors, we need to calculate the response kernel $K_{\mu\nu}$,
which is closely related to the response current density $J_{\mu}$ and the vector potential $A_{\nu}$ in terms of the liner response theory as
\cite{fukuyama69},
\begin{equation}\label{linres}
J_{\mu}({\bf q},\omega)=-\sum\limits_{\nu=1}^{3} K_{\mu\nu}({\bf q},\omega)A_{\nu}({\bf q},\omega),
\end{equation}
with the Greek indices label the axes of the Cartesian coordinate system. This doping and temperature dependence of the response kernel
(\ref{linres}) can be separated into two parts as
$K_{\mu\nu}({\bf q},\omega)=K^{({\rm d})}_{\mu\nu}({\bf q},\omega)+K^{({\rm p})}_{\mu\nu} ({\bf q}, \omega)$, where
$K^{({\rm d})}_{\mu\nu}({\bf q},\omega)$ and $K^{({\rm p} )}_{\mu\nu}({\bf q},\omega)$ are the corresponding diamagnetic and paramagnetic parts,
respectively. In this case, we \cite{feng10} have discussed the doping and temperature dependence of the electromagnetic response in the hole-doped
cuprate superconductors, and results show that the the electromagnetic response consists of two parts, the diamagnetic current and the paramagnetic
current, which exactly cancels the diamagnetic term in the normal state, and then the Meissner effect is obtained for all the temperature
$T\leq T_{c}$ throughout the SC dome. Following our previous discussions for the hole-doped case \cite{feng10}, the diamagnetic and paramagnetic
parts of the response kernel $K^{({\rm d})}_{\mu\nu}({\bf q},\omega)$ and $K^{({\rm p} )}_{\mu\nu}({\bf q},\omega)$ in the electron-doped cuprate
superconductors can be obtained in the the static limit as,
\begin{widetext}
\begin{subequations}\label{allkernel}
\begin{eqnarray}
K_{\mu\nu}^{(\rm{d})}({\bf q},0)&=&-{4e^{2}\over\hbar^{2}}(\chi_{1}\phi_{1}t-2\chi_{2}\phi_{2}t')\delta_{\mu\nu}={1\over \lambda^{2}_{L}}
\delta_{\mu\nu},\label{diakernel} \\
K_{\mu\nu}^{(\rm{p})}({\bf q},0)&=&{1\over N}\sum\limits_{{\bf k}}{\bf \gamma}_{\mu}({\bf k}+{\bf q},{\bf k}) {\bf \gamma}^{*}_{\nu}
({\bf k}+{\bf q},{\bf k})[L_{1}({\bf k},{\bf q})+L_{2}({\bf k},{\bf q})]=K_{\mu\mu}^{(\rm{p})}({\bf q},0)\delta_{\mu\nu}, ~~~~~
\label{parakernel}
\end{eqnarray}
\end{subequations}
where the charge carrier particle-hole parameters $\phi_{1}=\langle a^{\dagger}_{i\sigma} a_{i+\hat{\eta}\sigma}\rangle$ and
$\phi_{2}=\langle a^{\dagger}_{i\sigma}a_{i+\hat{\tau}\sigma}\rangle$, $\lambda^{-2}_{L}=-4e^{2}(\chi_{1}\phi_{1}t- 2\chi_{2} \phi_{2}t')/\hbar^{2}$
is the London penetration depth, and now is doping and temperature dependent, while the bare current vertex
${\bf \gamma}_\mu({\bf k}+{\bf q},{\bf k})$,
\begin{eqnarray}
{\mathbf{\gamma}}_\mu(\vv{k}+\vv{q},\vv{k})= \left
\{\begin{array}{ll} -{2e\over\hbar}\, e^{{1\over 2}iq_{\mu}}
\{\sin(k_{\mu}+{1\over 2}q_{\mu})[\chi_{1}t-2\chi_{2}t'
\sum\limits_{\nu\neq\mu}\cos({1\over 2}q_{\nu})\cos(
k_{\nu}+{1\over 2}q_{\nu})]\\
-i(2\chi_{2}t')\cos(k_{\mu}+{1\over 2}q_{\mu})
\sum\limits_{\nu\neq\mu}\sin q_{\nu}\sin(k_{\nu}+{1\over 2}q_{\nu})
\}\tau_0
  & {\rm{for}}\ \mu\neq 0,~~~~\\
{e\over 2}\tau_3  & {\rm{for}}\ \mu=0,\\
\end{array}\right.\label{barevertex}
\end{eqnarray}
with the functions $L_{1}({\bf k},{\bf q})$ and $L_{2}({\bf k},{\bf q})$ are given by,
\begin{subequations}\label{lfunctions}
\begin{eqnarray}
L_{1}({\bf k},{\bf q})&=&Z^2_{\rm aF}\left(1+{\bar{\xi}_{{\rm a}{\bf k}}\bar{\xi}_{{\rm a}{\bf k}+{\bf q}}+\bar{\Delta}_{\rm aZ}({\bf k})
\bar{\Delta}_{\rm aZ}({\bf k}+{\bf q})\over E_{{\rm a}{\bf k}}E_{{\rm a}{{\bf k}+{\bf q}}}}\right){n_{F}(E_{{\rm a} {\bf k}})
-n_{F}(E_{{\rm a}{{\bf k}+{\bf q}}})\over E_{{\rm a}{\bf k}}-E_{{\rm a}{{\bf k}+{\bf q}}}}, \\
L_{2}({\bf k},{\bf q})&=&Z^2_{\rm aF}\left(1-{\bar{\xi}_{{\rm a}{\bf k}}\bar{\xi}_{{\rm a}{\bf k}+{\bf q}}+\bar{\Delta}_{\rm aZ}({\bf k})
\bar{\Delta}_{\rm aZ}({\bf k}+{\bf q})\over E_{{\rm a}{\bf k}}E_{{\rm a}{{\bf k}+{\bf q}}}}\right){n_{F}(E_{{\rm a} {\bf k}})+
n_{F}(E_{{\rm a}{{\bf k}+{\bf q}}})-1\over E_{{\rm a}{\bf k}}+E_{{\rm a}{{\bf k}+{\bf q}}}}. ~~~~~
\end{eqnarray}
\end{subequations}
\end{widetext}
As in the hole-doped case \cite{feng10}, it is easy to show that in the long wavelength limit, i.e.,
$|{\bf q}|\to 0$, $K_{yy}^{({\rm p})}({\bf q}\to 0,0)=0$ at zero temperature ($T=0$). In this case, the long wavelength electromagnetic
response is determined by the diamagnetic part of the kernel only. On the other hand, at the SC transition temperature ($T=T_{c}$),
$K_{yy}^{({\rm p} )}({\bf q}\to 0,0)=-(1/\lambda^{2}_{L})$, which exactly cancels the diamagnetic part of the response kernel (\ref{diakernel}),
and then the Meissner effect with an external magnetic field is obtained for all $T\leq T_{c}$ throughout the SC dome.

However, as in the hole-doped case \cite{feng10}, the result we have obtained the response kernel of the electron-doped cuprate superconductors
in Eq. (\ref{allkernel}) can not be used for a direct comparison with the corresponding experimental data of the
electron-doped cuprate superconductors because the response kernel derived within the linear response theory describes the response of an
\emph{infinite} system, whereas in the problem of the penetration of the field and the system has a surface, i.e., it occupies a half-space $x>0$.
In such problems, it is necessary to impose boundary conditions for charge carriers. This can be done within the simplest specular reflection model
\cite{Abrikosov88} with a two-dimensional geometry of the SC plane. Taking into account the two-dimensional geometry of the electron-doped cuprate
superconductors within the specular reflection model \cite{Abrikosov88}, we can obtain explicitly the magnetic field penetration depth as
\cite{feng10},
\begin{eqnarray}\label{lambda}
\lambda(T)={1\over B}\int\limits_{0}^{\infty}h_{z}(x)\,{\rm{d}}x={2\over\pi}\int\limits_{0}^{\infty}{\rm{d}q_x\over\mu_{0}
K_{yy}(q_{x},0,0)+q_{x}^{2}},~~~~
\end{eqnarray}
which therefore reflects the measurably electromagnetic response in the electron-doped cuprate superconductors. It has been shown that there is a
similar strength of the magnetic interaction $J\approx 0.1\sim 0.13$eV for both hole-doped and electron-doped cuprate superconductors
\cite{hybertson90,gooding94,pavarini01}. Although the values of $t$ and $t'$ in the Hamiltonian (\ref{t-jmodel}) are believed to vary somewhat from
compound to compound, the numerical calculations \cite{hybertson90,gooding94,pavarini01} have extracted the range of these parameters for the
electron-doped cuprate superconductors as $t/J\approx -2.5\sim -3$ and $t'/t\approx 0.2\sim 0.3$. In this case, as a qualitative discussion in this
paper, the commonly used parameters are chosen as $t/J=-2.5$, $t'/t=0.3$, and $J=0.13$eV$\approx 1500$K. Furthermore, for the convenience in
the following discussions, we introduce a characteristic length scale $a_{0}=\sqrt{\hbar^{2}a/\mu_{0}e^{2}J}$. Using the lattice parameter
$a\approx 0.396$nm \cite{lian05} for the electron-doped cuprate superconductor Pr$_{2-x}$Ce$_{x}$CuO$_{4-\delta}$, this characteristic length is
obtain as $a_{0}\approx 80.88$nm.

\begin{figure}[h!]
\includegraphics[scale=0.48]{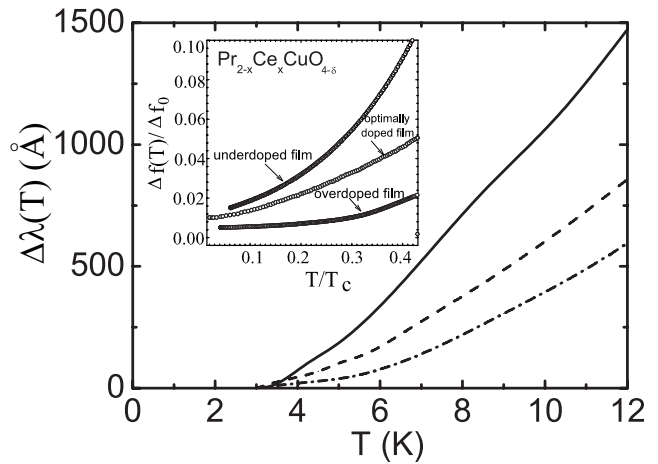}
\caption{The temperature dependence of the magnetic field penetration depth $\Delta\lambda(T)$ for $\delta=0.13$ (solid line), $\delta=0.15$
(dashed line), and $\delta=0.20$ (dash-dotted line) with $t/J=-2.5$, $t'/t=0.3$, and $J=1500$K. Inset: the corresponding frequency shift of the
resonator for Pr$_{2-x}$Ce$_{x}$CuO$_{4-\delta}$ taken from Ref. \onlinecite{Snezhko04} .
\label{fig1}}
\end{figure}

We are now ready to discuss the doping and temperature dependence of the electromagnetic response in the electron-doped cuprate superconductors.
Firstly, we discuss two limited cases of $T=0$ and $T=T_{c}$, respectively. At $T=0$, the magnetic field penetration depths are obtained as
$\lambda(0)\approx 366.61$nm, $\lambda(0)\approx 344.62$nm, and $\lambda(0)\approx 308.31$nm for the underdoping $\delta=0.13$, the optimal doping
$\delta=0.15$, and the overdoping $\delta=0.20$, respectively, which are consistent with the values of the magnetic field penetration depth
$\lambda\approx 115$nm $\sim 470$nm observed for different families of the electron-doped cuprate superconductors at different doping concentrations
\cite{Wu93,Andreone94,Kim03,Skinta02,Cooper96,Snezhko04,Kokales00,Prozorov00}. On the other hand, when $T=T_{c}$, we find that the kernel of the
response function $K_{\mu\nu}(\vv{q}\to 0,0)|_{T=T{c}}=0$, this leads to the magnetic field penetration depth $\lambda(T_{c})=\infty$, reflecting
that in the normal state, the external magnetic field can penetrate through the main body of the system, therefore there is no the Meissner effect
in the normal state. To analyze the evolution of the electromagnetic response with temperature, we have performed a calculation for the magnetic
field penetration depth (\ref{lambda}) with different temperatures, and the results of the $\Delta\lambda(T)=\lambda(T)-\lambda(0)$ as a function
of temperature $T$ for $\delta=0.13$, (solid line), $\delta=0.15$ (dashed line), and $\delta=0.20$ (dash-dotted line) are plotted in Fig. \ref{fig1}
in comparison with the corresponding frequency shift $\Delta f(T)$ of the resonator \cite{Snezhko04} for Pr$_{2-x}$Ce$_{x}$CuO$_{4-\delta}$ (inset).
This frequency shift $\Delta f(T)$ is proportional to the change of the magnetic field penetration depth $\Delta\lambda(T)$, i.e.,
$\Delta f(T)=f(T)-f(0)=G[\lambda(T)-\lambda(0)]$, where $G$ is a geometrical factor that depends upon the sample shape and volume, as well as
the coil geometry \cite{Snezhko04}. As in the hole-doped case \cite{kamal98,feng10}, our results show nearly linear characteristics of the magnetic
field penetration depth, except for the extremely low temperatures ($T<8$K), where a strong deviation from the linear characteristics (a nonlinear
effect) appears, which are in qualitative agreement with the experimental data of the electron-doped cuprate superconductors
\cite{Snezhko04,Kokales00,Prozorov00}.

\begin{figure}[h!]
\includegraphics[scale=0.43]{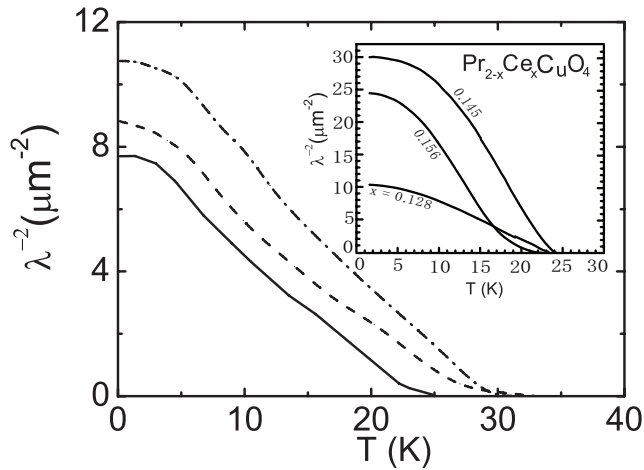}
\caption{The temperature dependence of the superfluid density for $\delta=0.13$ (solid line), $\delta=0.15$ (dashed line), $\delta=0.20$
(dash-dotted line) with $t/J=-2.5$, $t'/t=0.3$, and $J=1500$K. Inset: the corresponding experimental data of Pr$_{2-x}$Ce$_{x}$CuO$_4$
taken from Ref. \onlinecite{Skinta02}.
\label{fig2}}
\end{figure}

Now we turn to discuss the doping and temperature dependence of the superfluid density $\rho_{\rm s}(T)$, which is proportional to the squared
amplitude of the macroscopic wave function, and therefore describes the SC charge carriers. With the help of the magnetic field penetration
depth in Eq. (\ref{lambda}), the superfluid density of the electron-doped cuprate superconductors can be expressed as,
\begin{eqnarray}\label{superfluid}
\rho_{\rm s}(T)\equiv \lambda^{-2}(T).
\end{eqnarray}
Since the magnetic field penetration depth $\lambda(T_{c})=\infty$ at $T=T_{c}$ as mentioned above, this leads to the superfluid density
$\rho_{\rm s}(T_{c})=0$. For a better understanding of the evolution of the superfluid density with temperature, we have further performed a
calculation for the superfluid density (\ref{superfluid}) with different temperatures, and the results of $\rho_{\rm s}(T)$ as a function of
temperature for $\delta=0.13$ (solid line, $T_{c}=28$K), $\delta=0.15$ (dashed line, $T_{c}=34$K), and $\delta=0.20$ (dash-dotted line, $T_{c}=31$K)
are plotted in Fig. \ref{fig2} in comparison with the corresponding experimental results \cite{Skinta02} of Pr$_{2-x}$Ce$_{x}$CuO$_4$ (inset).
Our results show clearly that the superfluid density $\rho_{\rm s}(T)$ exhibits a linear variation with temperatures, however, in correspondence
with the nonlinear temperature dependence of the magnetic field penetration depth at the extremely low temperatures ($T<8$K) as shown in Fig.
\ref{fig1}, $\rho_{\rm s}(T)$ also crosses over to a nonlinear temperature behavior at the extremely low temperatures, which also are qualitatively
consistent with the experimental data of the electron-doped cuprate superconductors \cite{Skinta02,Cooper96,Snezhko04,Kokales00,Prozorov00}.

The essential physics of the doping and temperature dependence of the electromagnetic response in the electron-doped cuprate superconductors is the
same as in the hole-doped case \cite{feng10} except the nonmonotonic d-wave gap (\ref{gap}). Our results indicate that although the nonmonotonic
d-wave SC gap (\ref{gap}) modulates the renormalized charge carrier quasiparticle spectrum in the electron-doped cuprate superconductors, it does not
effect the overall global feature of the doping and temperature dependent magnetic field penetration depth (then the superfluid density). A crossover
from the linear temperature dependence at low temperatures to a nonlinear one at the extremely low temperatures in the magnetic field penetration
depth (then the superfluid density) is a basic consequence of the d-wave SC gap. However, although the momentum dependence of the SC gap (\ref{gap})
in the electron-doped cuprate superconductors obviously deviates from the monotonic d-wave SC gap \cite{matsui05}, it is basically consistent with
the d-wave symmetry. This is why the behavior of the electromagnetic response in the electron-doped cuprate superconductors is similar to that
observed in the hole-doped cuprate superconductors with a monotonic d-wave gap.

In summary, within the framework of the kinetic energy driven SC mechanism, we have studied the doping and temperature dependence of the
electromagnetic response in the electron-doped cuprate superconductors. Our results show that although there is an electron-hole asymmetry in the
phase diagram, the electromagnetic response in the electron-doped cuprate superconductors is similar to that observed in the hole-doped cuprate
superconductors. The superfluid density depends linearly on temperature at the low temperatures, except for the strong deviation from the linear
characteristics at the extremely low temperatures.

\acknowledgments

This work was supported by the National Natural Science Foundation of China under Grant No. 11074023, and the funds from the
Ministry of Science and Technology of China under Grant Nos. 2011CB921700 and 2012CB821403.

\end{document}